\documentclass[twocolumn,twoside,preprintnumbers,showkeys]{revtex4}
\usepackage{epsfig}
\usepackage{graphicx}
\usepackage{fancyhdr}
\usepackage{pslatex}

\begin{document}

\title{Distribution amplitudes of light mesons and photon in the instanton model
}\thanks{Based on the talks given at SPIN-Praha-2006, Prague, Czech, July 19 - 26, 2006
and ISMD-2006, Paraty, Rio de Janeiro, Brazil, September 02 - 08, 2006}
\author{Alexander E. Dorokhov}
\affiliation{Bogoliubov Laboratory of Theoretical Physics,\\
Joint Institute for Nuclear Research, 141980 Dubna, Russia}

\begin{abstract}
The leading- and higher-twist distribution amplitudes of pion, $\rho$-meson
and real and virtual photons are analyzed in the instanton liquid model.
\end{abstract}

\maketitle


\thispagestyle{fancy}

\setcounter{page}{1}

\section{Introduction}

Investigations of hard exclusive processes are essential for our
understanding of the internal quark-gluon dynamics of hadrons.
Theoretically, such studies are based on the assumption of factorization of
dynamics at long and short distances. The short-distance physics is well
elaborated by perturbative methods of QCD and depends on particular hard
subprocesses. The long-distance dynamics is essentially nonperturbative and
within the factorization formalism becomes parametrized in terms of hadronic
\emph{distribution amplitudes} (DAs). These nonperturbative quantities are
universal and are defined as vacuum-to-hadron matrix elements of particular
nonlocal light-cone quark or quark-gluon operators. The evolution of DAs at
sufficiently large virtuality $q^{2}$ is controlled by the renormalization
scale dependence of the quark bilinear operators within the QCD perturbation
theory. For leading-order DAs this dependence is governed by QCD evolution
equations. When the normalization scale goes to infinity the DAs reach an
ultraviolet fixed point and are uniquely determined by perturbative QCD.
However, the derivation of the DAs themselves at an initial scale $%
\mu_{0}^{2}$ from first principles is a nonperturbative problem and remains
a serious challenge.

Here we present the results \cite{Dorokhov:2002iu,Dorokhov:2006qm} of study
of the pion, $\rho $-meson and photon DAs in the leading and higher twists
at a low-momentum renormalization scale in the gauged non-local chiral quark
model \cite{Bowler:1994ir,Anikin:2000rq,Dorokhov:2003kf}
based on the instanton picture of QCD vacuum.

\section{Definitions and notations}

The distribution amplitudes of the mesons or the photon are defined via the
matrix elements of quark-antiquark bilinear operators taken between the
vacuum and the hadronic state $|h(q)\rangle $ of momentum $q$. It is assumed
that the quark and antiquark are separated by the distance $2z$ and the
light-like limit $z^{2}\rightarrow 0$ is taken at a fixed scalar product $%
q\cdot z$. We use the light-cone expansion of the matrix elements in order
to define the DAs \footnote{Our definitions of the photon and $\rho $-meson
DAs follow closely the works of Braun, Ball and coauthors \cite{Ball:1998sk,Ball:2002ps}.}
(only leading twist terms are presented)

\begin{equation}
\!\!\!\!\!\!\!\langle 0|\overline{d}(z)\gamma _{\mu }\gamma
_{5}[z,-z]u(-z)|\pi ^{+}(q)\rangle =i\sqrt{2}f_{\pi }q_{\mu
}\!\!\int_{0}^{1}\!\!\!dxe^{i\xi q\cdot z}\phi _{\pi }^{\mathrm{A}}(x),
\label{PiAV}
\end{equation}%
\begin{eqnarray}
&&\langle 0|\overline{q}(z)\sigma _{\mu \nu }[z,-z]q(-z)|\gamma ^{\lambda
}(q)\rangle =ie_{q}\left\langle 0\left\vert \overline{q}q\right\vert
0\right\rangle \chi _{\mathrm{m}}\cdot  \label{AT} \\
&&\cdot f_{\perp \gamma }^{t}\left( q^{2}\right) \left( e_{\mu }^{(\lambda
)}q_{\nu }-q_{\mu }e_{\nu }^{(\lambda )}\right) \int_{0}^{1}dxe^{i\xi q\cdot
z}\phi _{\perp \gamma }(x,q^{2}),  \nonumber
\end{eqnarray}%
\qquad
\begin{eqnarray}
&&\langle 0|\overline{q}(z)\gamma _{\mu }[z,-z]q(-z)|\gamma ^{\lambda
}(q)\rangle =e_{q}f_{3\gamma }f_{\parallel \gamma }^{v}\left( q^{2}\right)
q_{\mu }\cdot  \label{AV} \\
&&\cdot \frac{e^{(\lambda )}\cdot z}{q\cdot z}\int_{0}^{1}dxe^{i\xi q\cdot
z}\phi _{\parallel \gamma }(x,q^{2}),  \nonumber
\end{eqnarray}%
where $f_{\pi }$ is the pion decay constant, $\left\langle 0\left\vert
\overline{q}q\right\vert 0\right\rangle $ is the quark condensate, $\chi _{%
\mathrm{m}}$ is the magnetic susceptibility of the quark condensate, and $%
f_{3\gamma }$ is related to the first moment of the magnetic susceptibility.
The symbol $[-z,z]$ in the matrix elements denotes the path-ordered gauge
link (Wilson line) for the gluon fields between the points $-z$ and $z$. The
integration variable $x$ corresponds to the momentum fraction carried by the
quark and $\xi =2x-1$ for the short-hand notation. For a real photon, due to
condition $e^{(\lambda )}\cdot z=0$, the structure corresponding to $\phi
_{\parallel \gamma }$ decouples. The DAs $\phi _{\perp \rho }(x)$ and $\phi
_{\parallel \rho }(x)$ for the $\rho $-meson state $|\rho ^{\lambda
}(q)\rangle $ are defined in analogy with photon case (\ref{AT}) and (\ref%
{AV}) with mass-shell condition $q^{2}=-M_{\rho }^{2}$.

\section{Instanton-motivated nonlocal chiral quark model}

In the one loop approximation the quark model evaluation of the distribution
function $\phi _{h,J}(x)$ of hadron $h$ corresponding to projection $J$ is
given schematically as \cite{Dorokhov:2000gu}
\begin{equation}
N_{h,J}\phi _{h,J}(x)=-iN_{c}\int d\tilde{k}\delta (k\cdot n-x)\mathrm{Tr}%
[\Gamma _{J}S(k)\Gamma _{h}S(k-q)],  \label{phDA}
\end{equation}%
where the quark propagator has the form
\begin{equation}
S(p)=\frac{1}{\widehat{p}-M(p)+i\varepsilon },\qquad M(p)=M_{0}f^{2}(p^{2}),
\label{prop}
\end{equation}%
with the dynamical quark mass $M(p)$ expressed via the function $f\left(
p\right) $ defining the nonlocal properties of the QCD vacuum \cite{Dorokhov:1997iv}.
$\Gamma _{h}$ are the vertices defining the hadron state
\begin{eqnarray}
\Gamma _{\pi }(k,k^{\prime }) &=&\gamma _{5}f\left( k\right) f\left(
k^{\prime }\right) ,\quad \Gamma _{\rho }^{\mu }(k,k^{\prime })=\gamma _{\mu
}^{\perp }f_{V}\left( k\right) f_{V}\left( k^{\prime }\right) ,\quad
\nonumber \\
\Gamma _{\gamma }^{\mu }(k,k^{\prime }) &=&\gamma _{\mu }-(k+k^{\prime
})_{\mu }M_{k,k^{\prime }}^{(1)},  \label{vert1}
\end{eqnarray}%
and $\Gamma _{J}$\ is the projection operator corresponding to a definite twist. Here and
below, the notation%
\[
M^{(1)}\left( k,k^{\prime }\right) =\frac{M\left( k\right) -M\left(
k^{\prime }\right) }{k^{2}-k^{\prime 2}}
\]%
is used. The nonlocal functions are chosen in gaussian form
\begin{equation}
f(p)=f_{V}\left( p\right) =\exp \left( -\frac{p^{2}}{\Lambda ^{2}}\right) ,
\label{GaussFF}
\end{equation}%
with $p$ denoting the Euclidean momentum, resembling the fact that the
instanton field is convenient to take in the axial gauge. As the model
parameters we take the values fixed in \cite{Dorokhov:2004ze}
\begin{equation}
M_{0}=240~\mathrm{MeV},\;\;\Lambda =1110~\mathrm{MeV}.  \label{nlparam}
\end{equation}

\begin{figure}[th]
\hspace*{0.1cm} 
\vspace*{0.5cm} \epsfxsize=5.5cm \epsfysize=4.5cm
\centerline{\epsfbox
{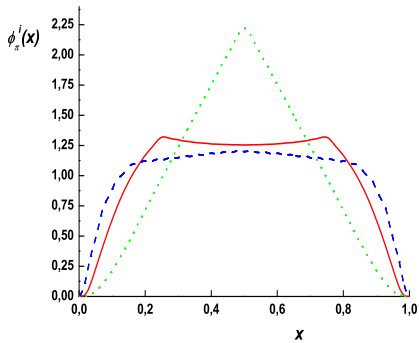}}
\caption[dummy0]{Pion distribution amplitudes: twist-2 axial-vector (solid
line), twist-3 pseudoscalar (short-dashed) and tensor (dotted) projections,
given at the quark model scale.}
\label{PiDA}
\end{figure}
\hspace*{0.5cm}
\begin{figure}[tbp]
\vspace*{0.5cm} \epsfxsize=5.5cm \epsfysize=4.5cm
\centerline{\epsfbox
{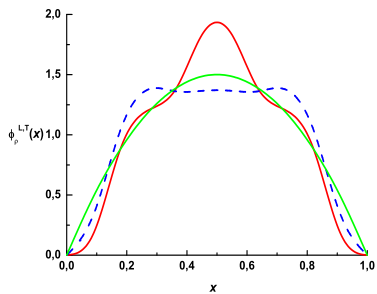}} 
\caption[dummy0]{$\protect\rho $-meson twist-2 distribution amplitudes:
transverse (solid line) and longitudinal (dashed) projections. The third
line is distribution amplitude at asymptotic scale.}
\label{RhoDA}
\end{figure}

The distribution amplitudes of the pion and the real photon calculated in
the instanton model in the chiral limit may be cast in a closed form. It is
convenient to introduce notations for the integration variables ($\overline{x%
}=1-x$)%
\begin{eqnarray*}
u_{+} &=&u-i\lambda x,\quad u_{-}=u+i\lambda \overline{x},\quad M_{\pm
}=M\left( u_{\pm }\right) ,\quad  \\
D(u) &=&u+M^{2}(u),\quad D_{\pm }=D\left( u_{\pm }\right) .
\end{eqnarray*}%
Then one gets the expressions%
\begin{equation}
\phi _{\pi }^{A}\left( x\right) =\frac{1}{f_{\pi }^{2}}\frac{N_{c}}{4\pi ^{2}%
}\int_{0}^{\infty }du\int_{-\infty }^{\infty }\frac{d\lambda }{2\pi }\frac{%
f_{+}f_{-}}{D_{+}D_{-}}\left( xM_{-}+\overline{x}M_{+}\right) ,
\label{PhiPiA}
\end{equation}%
\begin{eqnarray}
&&\phi _{\perp \gamma }\left( x,q^{2}=0\right) =\frac{1}{\left\vert \langle
\bar{q}q\rangle \right\vert \chi _{\mathrm{m}}}\frac{N_{c}}{4\pi ^{2}}\left[
\Theta (\overline{x}x)\int_{0}^{\infty }du\frac{M\left( u\right) }{D\left(
u\right) }-\right.   \nonumber \\
&&\left. -\int_{0}^{\infty }du\int_{-\infty }^{\infty }\frac{d\lambda }{2\pi
}\frac{M_{+}M_{-}}{D_{+}D_{-}}M^{\left( 1\right) }\left( u_{+},u_{-}\right) %
\right] .  \label{PhiTt}
\end{eqnarray}%
\begin{equation}
\phi _{\parallel \gamma }\left( x,q^{2}=0\right) =\Theta (\overline{x}x).
\label{wb1}
\end{equation}%
The DAs are scale dependent quantities and the above expressions correspond
to the low momentum scale $\mu _{0}$ typical for the instanton model. For
the instanton model it is estimated as $\mu _{0}=530$ MeV \cite%
{Dorokhov:2005pg}.  The parameters
entering normalization coefficients are given by
\begin{equation}
\left. \left\langle 0\left\vert \overline{q}q\right\vert 0\right\rangle ^{%
\mathrm{inst}}\right\vert _{1\mathrm{GeV}}=-(0.24~\mathrm{GeV)}^{3},\quad
\left. \chi _{\mathrm{m}}^{\mathrm{inst}}\right\vert _{1\mathrm{GeV}%
}=2.73\quad \mathrm{GeV}^{-2}.  \nonumber
\end{equation}

\begin{figure}[th]
\hspace*{0.1cm} \vspace*{0.5cm} \epsfxsize=5.5cm \epsfysize=4.5cm
\centerline{\epsfbox
{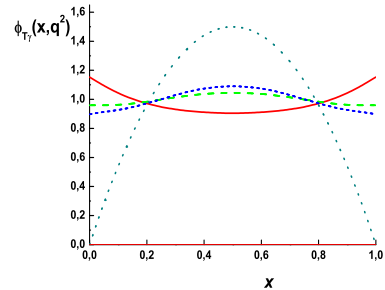}}
\caption[dummy0]{Dependence of the twist-2 tensor component of the photon DA
on transverse momentum squared ($q^{2}=0.25$ GeV$^{2}$ solid line, $q^{2}=0$
GeV$^{2}$ dashed line, $q^{2}=-0.09$ GeV$^{2}$ short-dashed line, asymptotic
DA - dotted line) given at the quark model scale.}
\label{TensorQ}
\end{figure}
\hspace*{0.5cm}
\begin{figure}[tbp]
\vspace*{0.5cm} \epsfxsize=5.5cm \epsfysize=4.5cm
\centerline{\epsfbox
{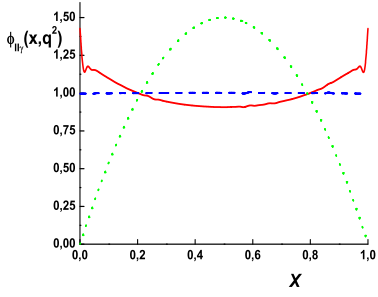}}
\caption[dummy0]{Same as Fig. \protect\ref{TensorQ} for the twist-2 vector
component of the photon DA.}
\label{FVii}
\end{figure}

The results of calculations are shown in Figs. 1-4. They correspond to low
momentum scale $\mu _{0}$ and need to be evolved to higher momenta scale in
order to compare with experimentally available information.
\begin{figure}[tbp]
\vspace*{0.5cm} \epsfxsize=5.5cm \epsfysize=4.5cm
\centerline{\epsfbox
{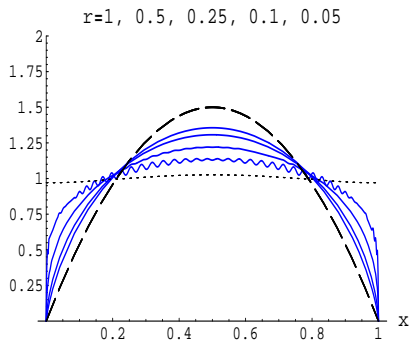}}
\caption[dummy0]{The LO ERBL evolution of the nonlocal model predictions for
the leading-twist \emph{tensor} projection of the real photon DA $\protect%
\phi _{\perp \protect\gamma }^{(t)}(x,q^{2})$. The dashed lines show the
asymptotic DA, $6x(1-x)$. Initial conditions, indicated by dotted lines, are
evaluated in the nonlocal quark model at the initial scale $\protect\mu ^{%
\mathrm{inst}}=530$~MeV. The solid lines correspond to evolved DA'a at
scales $Q=1$, $2.4$, $10$, and $1000$~GeV. The corresponding values of the
evolution ratio $r$ are given in the figures.}
\label{Evo}
\end{figure}
The DA at asymptotic scales $\mu _{\mathrm{as}}=\infty $ is also presented.

\section{Discussion}

Within the instanton model the leading twist pion
DA has been found in \cite{Esaibegian:1989uj,Dorokhov:1991nj,Petrov:1998kg}. The twist-3 and
twist-4 DAs have been found respectively in \cite{Esaibegian:1989uj} and \cite{Dorokhov:2002iu}.
The photon DAs up to twist-4 expansion have been found in \cite{Dorokhov:2006qm}.
Recently there have been published some papers
\cite{Noguera:2005cc,Nam:2006au,Nam:2006sx,Yu:15jd} where in nonlocal models similar
to considered above
the distribution functions for pion and photon were treated inconsistently.
Indeed, the typical expression (\ref{phDA}) defining the distribution functions has two
parts. One is coming from soft hadronic vertices (\ref{vert1}) and another refers to the operators
of definite twist which are responsible for the power corrections in the hard subprocess. It is well known
that, for example, the leading twist operators entering the pion distribution amplitude and
distribution function are given by $\gamma_\mu\gamma_5$ and $\gamma_\mu$, correspondingly.
Nevertheless, the authors of \cite{Noguera:2005cc,Nam:2006au,Nam:2006sx} included additional
terms proportional to $(k+k')_\mu$, where $k$ and $k'$ are incoming and outgoing momenta of a
quark. These additions in \cite{Noguera:2005cc,Nam:2006au,Nam:2006sx}
modify the known results of the leading twist calculations.
However, it is evident, see for example \cite{Dorokhov:2002iu}, that these additional terms
contribute to twist-4 distributions and do not touch the leading twist-2 distributions.
From the other side, the full vertex including the local and nonlocal pieces is important
in the soft hadronic part and has not to be
neglected as it was done in \cite{Petrov:1998kg,Yu:15jd} considering the photon distribution
amplitudes.

\section{Conclusions}

The instanton model of QCD vacuum is realistic tool to get
nonperturbative properties of hadrons in terms of parameters characterizing
the vacuum. All hadron DAs are suppressed at the bound of kinematical
interval due to localized wave function of hadrons, while photon DAs are not
zero there. By applying the QCD evolution the photon DAs become immediately
zero at the edge points of $x$-interval. Nevertheless, the photon DAs are
always wider than asymptotic distribution. The first experimental results on
direct measurements of pion and photon DAs are discussed in \cite%
{Aitala:2000hc,Ukleja:2004yt}.

\noindent\textbf{Acknowledgements}

The author thanks organizers for very fruitful meeting and the Russian
Foundation for Basic Research projects No. 04-02-16445, Scient. School grant
4476.2006.2 and the JINR Heisenberg-Landau program for partial support.

\end{document}